\documentclass[preprint,12pt]{elsarticle}

\usepackage{amsmath,amssymb,amsfonts,mathrsfs}
\usepackage{mathtools}
\DeclarePairedDelimiter{\ceil}{\lceil}{\rceil}
\usepackage[inline]{enumitem}
\usepackage{amsthm}
\usepackage{subcaption}
\newtheorem{theorem}{Theorem}

\journal{Physica}

\begin{document}

\begin{frontmatter}

\title{Fragmentation from group interactions: A higher-order adaptive voter model}

\author{Nikos Papanikolaou$^{a}$, Renaud Lambiotte$^{b}$, Giacomo Vaccario$^{c}$}

\affiliation{
organization={Institute of Theoretical Physics IV, University of Stuttgart},%
            country={Germany}
            }

\affiliation{organization={Mathematical Institute,
University of Oxford},%
            country={United Kingdom}}
\affiliation{
organization={Chair of Systems Design, ETH Zurich},%
            country={Switzerland}
            }

\begin{abstract}
The adaptive voter model allows for studying the interplay between homophily, the tendency of like-minded individuals to attract each other, and social influence, the tendency for connected individuals to influence each other. 
However, it relies on graphs, and thus, it only considers pairwise interactions. 
We develop a minimal extension of the adaptive voter model to hypergraphs to study the interactions of groups of arbitrary sizes using a threshold parameter. 
We study $S$-uniform hypergraphs as initial configurations. 
With numerical simulations, we find new phenomena not found in the counterpart pairwise models, such as the formation of bands in the magnetization and the lack of an equilibrium state. 
Finally, we develop an analytical model using a sparse hypergraph approximation that accurately predicts the bands' boundaries and height. 
\end{abstract}


\end{frontmatter}

\section{Introduction}
How collective phenomena can be explained from their micro-constituents is at the core of many disciplines.
For example, in statistical mechanics, the Lenz-Ising model explains the spontaneous magnetization of materials by considering the local interactions among two adjacent atomic dipoles~\citep{brush1967history}. 
Similarly, in socio-physics, the adaptive voter model describes the emergence of consensus and fragmentation in social networks by modelling interactions among individuals~\citep{adaptivevotermodel}. 
In this model, each individual $i$ is characterized by a degree of freedom $s_i=\{0,1\}$ representing whether the individual is in favour ($s_i=1$) or against ($s_i=0$) a given issue.
Then, individuals are connected among each other and interact according to a simple rule: an individual can either adopt the opinion of a neighbour or drop this connection and create a new one with an individual having the same opinion. 
Despite the simplicity of this dynamics, it exhibits two totally different final states: consensus (i.e., all individuals have the same opinion) or fragmentation -- where the social network splits into two separate components with opposite opinions. 

We extend this type of models considering \textit{group interactions} and study how they affect fragmentation.
Group interactions are interactions that involve more than two individuals. 
In opinion dynamics, examples are group messaging, group discussions or emails with multiple recipients. 
Studies have shown that complex mechanisms based on group interactions are often required to describe the dynamics in a social group~\citep{iacopini2019simplicial,iacopini2022group,landry2023polarization, ferraz2023multistability}. 
Examples of such mechanisms are peer pressure~\citep{brown1986perceptions} and reinforcement~\citep{milgram1969note}.
Another mechanism is advanced by Social Impact Theory, stating that groups modulate the impact of a source on a target individual~\citep{latane1981social,holyst2000phase}.
Moreover, group interactions have been very relevant in diverse fields ranging from  physics~\citep{goban2018emergence, pelka2020chimera}, neural networks~\citep{yu2011higher}, and ecology~\citep{levine2017beyond}. 

To model group interactions, we use hyperedges~\citep{battiston2020networks}.
A hyperedge of size $k\geq 2$ represents a group interaction among $k$ individuals.
By combining hyperedges, we obtain a hypergraph.
This mathematical object is a powerful tool successfully used to study diverse group interactions, such as multi-protein interactions in cellular biology~\citep{klimm2021hypergraphs}, species interactions in theoretical and experimental ecology~\citep{bairey2016high,golubski2016ecological}, and
academic teams in co-authorship networks~\citep{taramasco2010academic}.

One can also use simplicial complexes to model group interactions~\citep{salnikov2018simplicial}.
Simplicial complexes are hypergraphs with additional constraints~\citep{bick2021higher}.
An important one for this discussion is that their hyperedges are closed under inclusion. 
This requirement means that all the individuals of a group are assumed to also interact with each other, pairwise or in small groups.
We instead use hyperedges to depict group interactions.
This choice allows describing arbitrarily large social groups whose members do not necessarily interact pairwise or through smaller groups with all other members. 

Our model is based on the adaptive voter model of~\citep{nk2022} and has the following dynamics.
At each time step, we choose a hyperedge $e$ and check its size $n_e$. 
If $n_e=2$, i.e., a pairwise interaction, then we apply the rules of the adaptive voter model~\citep{adaptivevotermodel}. 
If $n_e\geq 3$, either the \textit{influence} or \textit{split-merge} process occurs.
The influence process assumes that the minority in the group adopts the majority's opinion with a certain probability.
The split-merge process instead assumes that the minority splits from the group and merges with another group sharing the same majority opinion.
A threshold parameter $\gamma$ determines the critical size of the minority at which either the influence or split-merge process occurs.

The main differences from~\citep{nk2022} are two.
First,~\citep{nk2022} studies the system in a heterogeneous mean-field regime (HMF) where the \textit{group size distribution} was preserved.
Hence, they studied how group interactions affect the dynamics to total consensus, but fragmentation could not emerge as the final state.
Here, we instead explore how fragmentation is affected by group interactions. 
To this end, we consider a system far away from an HMF.
Second,~\citep{nk2022} considers that when a group splits, both subgroups merge into other groups.
We instead assume that only the minority group merges.
This change implies that we now preserve the number of groups over time; hence, the importance of groups also stays constant during the dynamics.

We study how fragmentation is affected by the threshold parameter $\gamma$ and the initial mean degree, i.e., the average number of groups to which each individual belongs.
In general, we find that fragmentation decreases with gamma (i.e., the importance of group influence) and initial mean degree (i.e., the system's connectivity).
Moreover, we find a striking difference compared to the adaptive voter model without group interactions.
We find fragmentation bands, i.e., equilibrium states with different degrees of fragmentation depending on $\gamma$.
As the threshold parameter varies, the transition between these states is discontinuous.
We also provide an analytic explanation for these bands and their discontinuity when the hypergraphs are sparse.

The remainder of this paper is divided into three sections.
In Sect.~\ref{sec:model}, we present our model's dynamics and define the observables. 
In Sect.~\ref{sec:results}, we present the results: the effects of group interactions on fragmentation (Subsect.~\ref{sec:fragmentation}), a comparison adaptive voter model without group interactions (Subsect.~\ref{sec:comparisonclassic}), and an analytic model under a sparsity approximation (Subsect.~\ref{sec:sparse}). 
Finally, in Sect.~\ref{sec:conclusion}, we summarize the results and discuss future work.

\section{Hypergraph Adaptive Voter Model}
\label{sec:model}
We model individuals as nodes on a hypergraph. 
Each node $i$ is described by a state variable $s_i(t)$ which represents its opinion at a specific time $t$ and can take value either $1$ or $0$.  
Let $N$ be the number of nodes.

\subsection{Initialization}
At the beginning of each simulation, nodes are assigned the opinion $1$ with probability $A_s$.
We generally set $A_s$ = 0.5, unless stated. 
Also, we initialize the system (at $t=0$) as an $S$-uniform hypergraph $\mathcal{H}_0$ given by $(V, E)$ where $V$ is the vertex set and $E$ is the edge set, containing only edges of size $S$.
For the $S$-uniform hypergraph, the mean degree is $\mathcal{H}_0$ is $\langle d(\mathcal{H}_0) \rangle=\frac{S~ n}{N}$ where $n$ is the number of edges in $E$.
In summary, the parameters for initializing the model are $N$, $A_s$, $n$, and $S$.

In the next section, we define the dynamics of the model.
For the dynamics, there are two parameters: a \textit{probability of rewiring} $p$ and a \textit{threshold parameter} $\gamma \in [0, 0.5]$. 

\subsection{Dynamics of the model}
\begin{figure}
    \centering
    \footnotesize
(a)\includegraphics[height=0.32\textwidth]{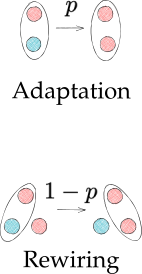}
\hfill
(b)\includegraphics[height=0.32\textwidth]{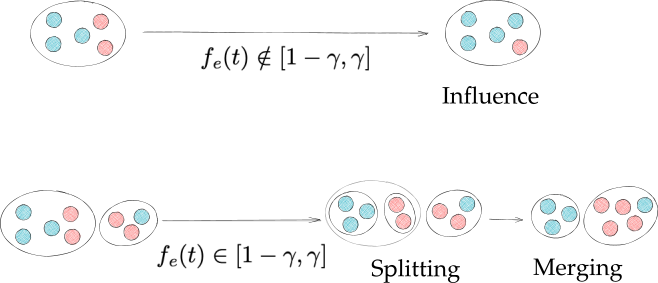}
    \caption{Schematic representation of the model dynamics. The red circles are nodes with opinion $0$ and the blue circles are nodes with opinion $1$. 
    Depending on whether it is a simple edge (a) or a hyperedge (b), we apply different rules. 
    In (a), we consider an active simple edge and have either rewiring or adaptation occurring with probability $1-p$ or $p$, respectively. 
    In 
    (b) we consider a hyperedge with fraction of nodes with opinion $1$ (red) equal to $\frac{3}{5}$.
    Influence occurs if $\frac{3}{5}>\gamma$, then each (red) node with minority opinion may change its opinion with probability $\frac{3}{5}$. 
    Else, splitting and merging occur: the active hyperedge splits into two. 
    The split edge with the minority opinion (red) is merged into a second edge with the same majority opinion chosen randomly from the rest of the network.}
\label{dynamics_image}
\end{figure}

We call an edge $e$ active (inactive) if the opinions of the nodes in $e$ are different (same). 
This definition applies to both simple edges and hyperedges. 
The fraction of nodes with opinions $1$ in an edge $e$ is denoted by:
\begin{equation}
    f_e(t) = \frac{1}{n_e} \sum_{i \in e} s_i(t)
\end{equation}
where $n_e$ is the size of the edge $e$, i.e., the number of nodes belonging to the group.
At each time step, we sample an edge $e$ from $E(t)$:
\begin{itemize}
    \item if $e$ is a simple edge (i.e. $n_e=2$) then 
    \begin{itemize}
        \item if $e$ is active then:
        \begin{itemize}
            \item with probability $p$, \textit{rewiring} occurs. This means that each node in $e$ rewires to a random node from the network with the same opinion.
            This process changes the edge set $E$.
            \item with probability $1-p$, \textit{adaptation} occurs. This means that one of the nodes is randomly chosen and it adopts the opinion of the other.
        \end{itemize}
        \item if $e$ is inactive then nothing happens.
    \end{itemize}
    \item if $e$ is a hyperedge (i.e., $n_e\geq 3$) then:
    \begin{itemize}
        \item if $f_e(t)\leq \gamma$ or $f_e(t) \geq 1-\gamma$ then \textit{influence} occurs.
        This means that each node with the minority opinion changes its opinion with probability proportional to $f_e(t)$ if the majority opinion is 1 and $1-f_e(t)$ otherwise.
        In case of a tie, the ``minority" opinion is chosen randomly. 
        This is an extension of the adaptation for group interactions.
        \item if $\gamma < f_e(t) < 1-\gamma$ then \textit{splitting and merging} occurs. 
        Splitting means that the hyperedge $e$ separates into two inactive edges with opposite opinions. 
        Merging means that the smaller split edge integrates with another edge randomly chosen from the hypergraph whose majority opinion is the same as in the split edge.\footnote{If the smaller split edge cannot find another edge sharing the same majority, then the larger edge is integrated with another edge. This allows to keep constant the number of hyperedges.} 
        This process changes the edge set $E$ and is an extension of the rewiring for group interactions: it models homophily at between groups.
    \end{itemize}
\end{itemize}
This procedure is repeated until equilibrium is reached. 
We define equilibrium as when all the edges have become inactive. 
When an edge $e$ has been selected at a timestep, we graphically depict its dynamics in Figure \ref{dynamics_image}. 

Based on this dynamics, hyperedges are treated differently from simple edges, i.e., their mechanisms are fundamentally different. 
The motivation behind this is that hyperedges model group interactions, while simple edges model pairwise interactions.
For the former, the concept of majority and minority emerges unlike for the latter. 
The presence of a majority and minority can create biases towards one opinion. 
In our model, we focus on the case in which the majority opinion is preferred above a certain threshold.
Note that the proposed dynamics for group interactions cannot be described as multiple pairwise interactions.
This impossibility to decompose the group interactions introduces new possible phenomena that could not be observed using models of pairwise interactions~\citep{noonan2021dynamics}.

\subsection{Difference from previous models}
\label{sec:differences}
Even though these dynamics are quite similar with those presented in the introduction, there are important differences. 
Firstly, we use hypergraphs instead of simplicial complexes. 
The main reason is that as explained in the previous sections hypergraphs are less constraining to model large groups, since it is not assumed that each subgroup or pair of individuals in a group are connected, as is the case in simplicial complexes (they are closed under inclusion).

Moreover, in our model, influence and splitting/merging take place deterministically depending on the threshold parameter and not on a probability $q$. 
This choice can be justified by the fact that we would expect influence to occur with different probabilities for different sizes of groups. 
Also, unlike for groups of size $3$, for big groups the fraction of minority opinions can take multiple values and we would like the large values to be treated differently than smaller ones.
For example, in a hyperedge with size $5$, an edge with only one node of opinion $0$ has different impact on the opinion dynamics than an edge with two nodes of opinion $0$.
This distinction would not be possible using a hyperparameter probability $q$ and it would be cumbersome to introduce one for every possible size and minority fraction. 
The concept of threshold in group interactions is also sociologically useful based on the threshold model. 

Another important difference is that at each timestep whole groups can be picked, unlike in the previous model where only simple edges were randomly chosen. In this way, we  decouple the group interactions from the pairwise interactions. This is because the dynamics of the selected hyperedge is determined by the threshold parameter and the fraction of minority opinions in that hyperedge independently of parameters describing pairwise interactions.

\subsection{Edge-based magnetization}

Similarly to the classical Voter Model, the quantities of interest are the time to reach equilibrium and the magnetization. 
The former is useful to investigate whether group interactions accelerate or delays the evolution of system to its equilibrium. 
The latter quantifies whether the system at equilibrium has reached consensus or its degree of fragmentation.

We distinguish two kinds of magnetization: edge-based magnetization and node-based magnetization. 
Both kinds of magnetization for finite systems can be used to distinguish whether total consensus or fragmentation occurs. 
If the magnetization at equilibrium is equal to 1 (-1) then the system has reached total consensus with opinion 1 (0). Otherwise it is fragmented. 
In formula, the node-based magnetization at time $t$ is defined as: 

\begin{equation}
  m(t) =
  \frac{\sum_{\substack{i\in V}}
        (2s_i(t)-1)}{N},
\label{m_node}
\end{equation}
where $s_i$ is the opinion of the node $i$. 
This means that the node-based magnetization is equal to the fraction of nodes with opinion 1 minus the fraction of nodes with opinion 0 in the node set $V$.
Edge-based magnetization at time $t$ sums through all the nodes of all the edges and is defined as:
\begin{equation}
  m(t) =
  \frac{1}{\sum_{
                  e\in E(t)}n_e} \sum_{\substack{i\in e\\
                  e\in E(t)}}(2s_i(t)-1)
\label{m_edge}
\end{equation}
where $s_i$ is the opinion of the node $i$. 
The edge-based magnetization is related to degree-weighted moments because the nodes with the greatest number of degrees contribute the most to the sum.

In our study, choosing the node-based over the edge-based magnetization makes negligible difference since the initial hypergraph is chosen to have a binomial degree distribution. 
Therefore, the standard deviation of the degree is relatively small, and hence, each node is treated equal in the edge-based magnetization. 
For this study, we chose the edge-based magnetization as it is usually preferred in the complex network literature.

\section{Results}
\label{sec:results}
\subsection{Fragmentation of adaptive voters on hypergraphs}
\label{sec:fragmentation}
\begin{figure}
    \centering
    \footnotesize
    \includegraphics[width=1\textwidth]{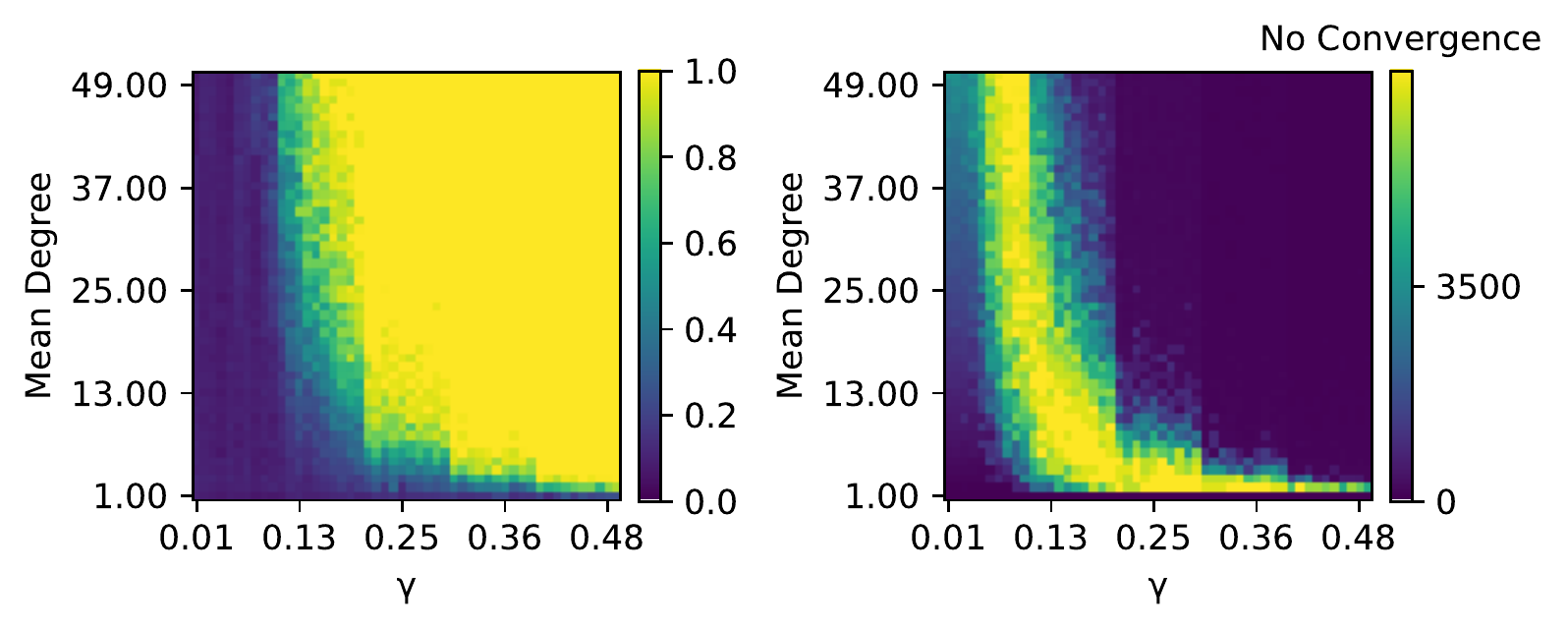}
    \caption{Colormaps of mean degree versus $\gamma$. (a) The color represents the average absolute magnetization.
    (b) The color represents the average time to convergence. The maximum number of simulated steps was chosen to be $7000$. If a trajectory does not converge then this number is assigned for computing the average. The parameters are: $p=0.55$, $N=100$, $S=10$, $A_S=0.5$. For each mean degree and $\gamma$, we simulate $20$ trajectories.}
    \label{fig:heatmap}
    \end{figure}
\subsubsection{Low values of $\gamma$ lead to fragmentation}
In Fig.~\ref{fig:heatmap} (a), we report the absolute edge magnetization in function of $\gamma$ and the mean initial degree of the nodes.
In general, we observe that we have consensus for high values of $\gamma$, i.e., when the dynamics is dominate by influence.
At lower values of $\gamma$, we instead have fragmentation, i.e., the final state is composed by groups containing nodes with both opinions.
This occurs because at low $\gamma$ it is more likely that an active group splits into two groups having opposite opinion.
By this, a fragmented state emerges and consensus is out of reach.

\subsubsection{High mean degrees lead to total consensus}
\label{sec:highmean}
In Fig.~\ref{fig:heatmap} (a), we also observe that the fragmentation depends on the initial mean degree.
The mean degree describes the average number of groups to which a node belongs.
We explore mean degree values ranging from one to 100.
We find that at low initial mean degree, we have more fragmentation, while at high mean degree less.
This result occurs as when decreasing the mean degree, a node belongs to very few groups (one or two).
If one of this group splits, it is very unlikely that its nodes will go through the influence dynamics and fragmentation will appear.
For high mean degree, nodes belong to more groups and hence, groups overlap.
Thanks to this overlap, the majority opinion of the system can propagate, and the systems can reach total consensus. 

\subsubsection{Time of convergence}
\label{sec:timeofconvergence}
In Fig.~\ref{fig:heatmap} (b), we report the time needed to converge to a stable state, i.e., a state without active groups.
We find that the time of convergence changes a lot depending on $\gamma$ and the initial mean degree.
In particular, at fixed initial mean degree, the time of convergence is not monotonous in $\gamma$.
It first increases with $\gamma$ until a threshold value $\gamma_t$ that depends on the mean degree.
Then there is a $\gamma$-range in which the time of convergence stays high, and finally it decreases again.

At very low $\gamma$, fragmentation is a stable state from which the system does not move, see Fig.~\ref{fig:heatmap} (a). 
This state is also quick to reach as almost every sampled group splits into two.
Hence, the time of convergence is of the order of the number of initial groups (Fig.~\ref{fig:heatmap} (b)).
When increasing $\gamma$, groups undergo the influence dynamics which tries to move the system towards consensus.
However, at this intermediate values of $\gamma$, influence is not strong enough and does not manage to push the system to consensus. 
The system gets instead trapped in $k$-orbits and never reaches a state without active groups. 

A simple example of $k$-orbit is when a node $i$ is in two groups with opposite majority opinion: a first group active and a second one inactive.
When influence acts on the active group and changes the opinion of the node $i$, the first group becomes inactive and the second one active.
This type of dynamics can repeat itself, locking the system on a 2-orbit.
Above $\gamma_t$ and the range with the $k$-orbits, consensus is the final equilibrium of the system, see Fig.~\ref{fig:heatmap} (a). 
The bigger is $\gamma$, the more likely influence occurs, and hence, the time to reach consensus decreases, see Fig.~\ref{fig:heatmap} (b).

\subsection{Comparison to the adaptive voter model without group interactions}
\label{sec:comparisonclassic}
\subsubsection{Predicting the final state}
In the adaptive voter model without group interactions by \citep{adaptivevotermodel}, the magnetization (i.e., fraction of ones, $m$) and the density of active edges (i.e., fraction of 0-1 edges, $\rho_e$) describe the phase transition between fragmentation and consensus. 
The magnetization denotes whether the system is in fragmentation or total consensus. 
The density of active edges denotes whether the system has reached equilibrium. 
In \citep{adaptivevotermodel}, the authors show that the density of active edges has a quadratic form (a concave parabola) in the magnetization during the system's evolution, i.e., $\rho_e(t)\sim -m(t)^2$. Hence, by fitting a parabola on the time sequence of ($m(t)$,$\rho_e(t)$), the intersections between the fitted parabola and the $x$-axis predict the final states that the system eventually reaches at $t \to \infty$.
Prompted by this result, we ask whether it applies also in presence of group interactions.

\begin{figure}
    \centering
    \includegraphics[width=0.6\textwidth]{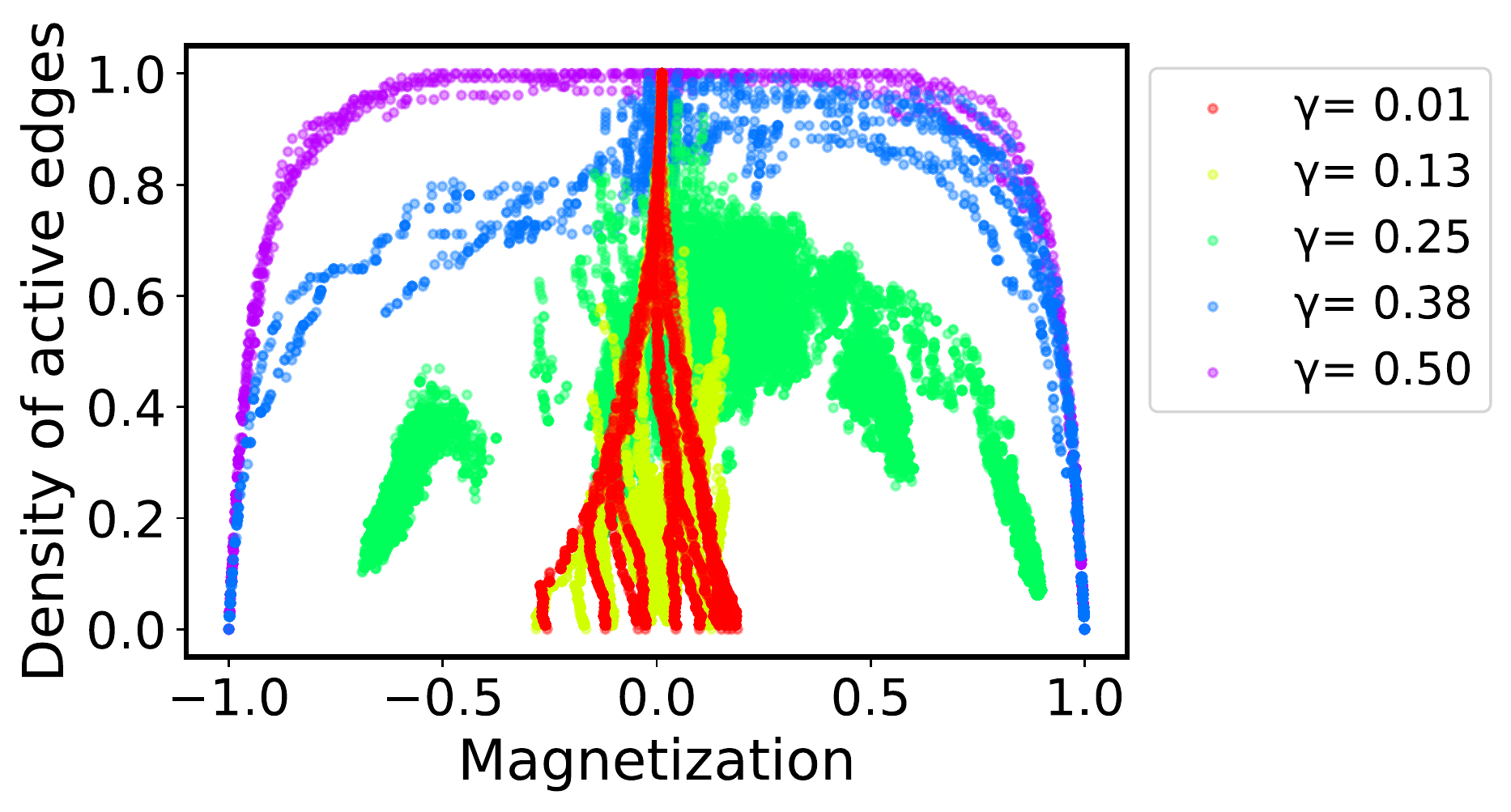}
    \caption{The edge-based magnetization for $10$ trajectories for $5$ values of threshold parameters $\gamma$ versus the density of active edges. The parameters are $p=0.55$, $N=500$, $S=32$, $n=125$ with mean degree, $\langle d(\mathcal{H}_0) \rangle=8$. The curves cannot be fitted with parabolas due to a congestion of trajectories around density of active edges equal to 1.}
    \label{fig:active_edges}
\end{figure}

In Fig. \ref{fig:active_edges}, we plot the density of active edges versus edge-based magnetization for different values of $\gamma$.
We find that parabolas are poorly fitted to the trajectories and the results of~\citep{adaptivevotermodel} do not generalize for the dynamics of the presented model.
The reason is that there is a congestion of trajectories at high values for the density of active edges.
This congestion happens because a group is active if there is at least one node with a different opinion. 
Therefore, in each group multiple nodes need to change their opinion to change the group from active to inactive.
Since we initialize nodes with random opinions and also group them at random, the density of active edges starts close to 1.
Then, there are large time periods during which the magnetization can change while the density of the active edges stays constant.
This process creates a congestion of trajectories at high values for the density of active edges.

\subsubsection{Multiple fragmented states}
\begin{figure}
    \centering
    \footnotesize
    (a)
  \includegraphics[width=0.45\textwidth]{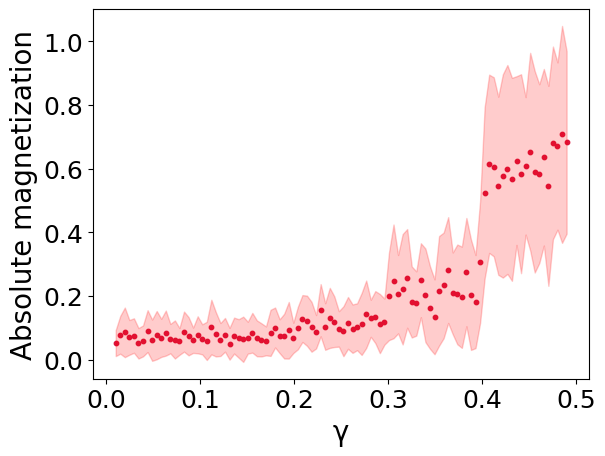}
    \hfill
  \includegraphics[width=0.45\textwidth]{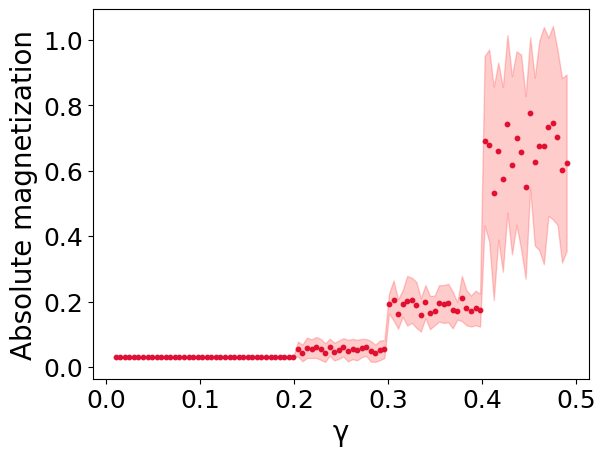}
\caption{
Absolute final \textit{edge-based} magnetization versus threshold parameter, $\gamma$. 
In (a), the merging and rewiring mechanism is active while in in (b), they are not. 
The initial configuration is an $S$-uniform hypergraph with edge size, $S=10$, number of nodes $N=100$, number of initial hyperedges $n=20$ (thus the mean degree is 2) and probability of rewiring $p=0.55$.  
Opinions $1$ and $0$ were initially assigned to the nodes with equal probability. The points and the shaded area are the mean and the standard deviation of the absolute magnetization for $20$ trajectories for each value of $\gamma$. 
In both (a) and (b), there are bands where the average absolute magnetization  stays constant and then increases abruptly. 
These bands are more noticeable in (b).}
\label{fig:edge_abs_gamma}
\end{figure}
In Fig~\ref{fig:edge_abs_gamma}(a), we show the absolute final edge-based  magnetization versus $\gamma$. 
We find the occurrence of bands, i.e., the absolute magnetization is approximately a step function between the initial value of the magnetization $2A_s-1$ and 1. 
In other words, we have \textit{multiple} fragmented states when varying the strength of influence.
This phenomenon is not observed in the adaptive voter model without group interactions which has one phase transition.  
Precisely, when increasing the probability to rewire active edges, the final possible states are \textit{only two}: a state with the fraction of minority equal to zero (total consensus) or a state with the fraction of minority equal to the initial (minority) fraction. 

The presence of bands with constant magnetization is only an effect of influence.
Indeed, they are more noticeable if there is no the merging and rewiring mechanism, but only influence and splitting (see Fig~\ref{fig:edge_abs_gamma}(b)). 
This means that merging and rewiring is unrelated to the existence of the bands. 
The effect of these two mechanisms is to smooth out the curve, especially for low values of $\gamma$ where they are more likely to occur. 

In Sect.~\ref{sec:sparse}, we derive the height (i.e., the absolute magnetization value) and the location (i.e., the $\gamma$ ranges) of these bands when the merging and rewiring process are switched off and the hypergraphs is sparse.

\begin{figure}
    \centering
    \footnotesize
    (a)
    \includegraphics[width=0.45\textwidth]{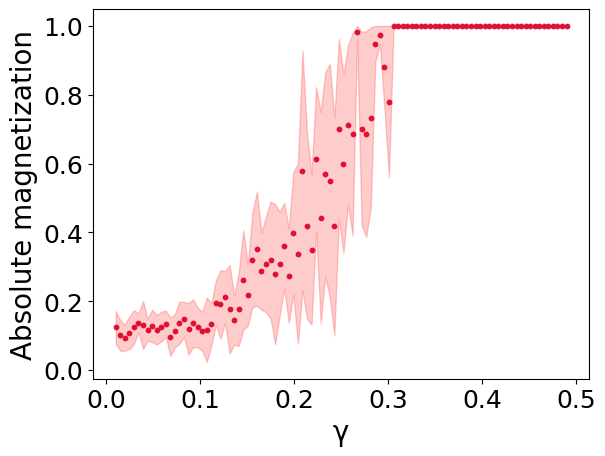}
    \hfill
    (b)
    \includegraphics[width=0.45\textwidth]{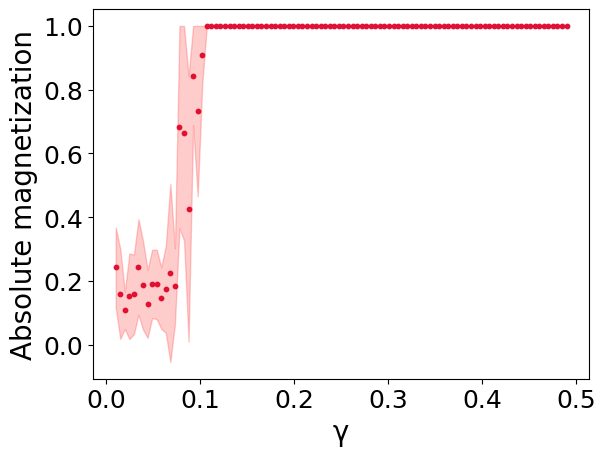}
    \caption{Absolute final edge-based magnetization versus $\gamma$. (a) Mean degree equal to $10$ ($N=100$,  $n=100$, $S=10$). (b) Mean degree equal to $100$ ($N=100$, $n=1000$, $S=100$). The points and the shaded area are the mean and the standard deviation of the trajectories respectively for 10 trajectories for each of the 100 values of $\gamma$ used. The other parameters are $p=0.55$ and $A_s=0.5$.}
    \label{fig:m_gamma_example}
\end{figure}

\subsubsection{Convergence to the adaptive voter model without group interactions}
For large number of groups $n$, the presented model looks similar to the adaptive voter model without group interactions. 
In Fig.~\ref{fig:m_gamma_example}, we plot the absolute magnetization vs $\gamma$ for mean degree $10$ and $100$. 
The absolute magnetization for most of the values of $\gamma$ is either equal to the $2A_s -1$ (with $A_s=0.55$) or equal to 1. 
This phenomenon occurs also in the classical adaptive voter model.

Note also that there is a small region of $\gamma$ values for which the absolute magnetization has intermediate values. 
When increasing the mean degree, the ``width'' of the bands decreases and the system asymptotically approaches a sharp transition. 

To better understand the boundaries of the bands and how the absolute magnetization changes, we now analytically study our model.
We consider the case where the mean degree is low since the bands are more prominent in this regime (see Fig.~\ref{fig:heatmap} (a)).

\subsection{Sparse hypergraph approximation}
\label{sec:sparse}

We develop an analytical expression to describe the boundaries and the height of the bands, i.e., we characterize the multiple fragmented states.
For this analysis, we assume that the hypergraph is sparse.
The sparsity assumption allows us to ignore the overlap between edges.
Also, we neglect the merging and rewiring mechanism since bands still exist without this mechanism.

\subsubsection{The boundaries of the bands}
In Appendix, we formally prove the existence of the bands and their boundaries with respect to the parameter $\gamma$. 
To do this, we calculate the master equation for $N(k,l,t)$ that is the number of edges with $k$ nodes of opinion 1 and size $l$ at time $t$. 
We recursively solve this equation and find that the boundaries of the bands are the rational numbers $\frac{k}{S}$ where $k=1,2,...,S$ such that $N(k,S,0)$ is larger than zero.

A heuristic proof is the following. 
Without loss of generality, let us assume that $A_s>0.5$.
Then, at $t=0$, the majority of edges of size $S$ that do not split will on average become inactive edges of opinion $1$ due to the influence mechanism. 
These edges increase the absolute edge-based magnetization and this increase is preserved in time by the sparsity of the hypergraph.
Precisely, the sparsity hypothesis implies that there is little overlap between edges, and hence, inactive edges stay inactive as they cannot be re-activated from nodes belonging to other edges.
On the other hand, if an initial hyperedge splits, it creates two inactive edges of opposite opinions. 
These inactive edges do not change the absolute edge-based magnetization and their state stays frozen.
Hence, the final absolute edge-magnetization depends on the \textit{initial} fraction of edges that is susceptible to influence or split.
A hyperedge is susceptible to influence or split depending on the value of $\gamma$ and its minority fraction which is a discontinuous value $\frac{k}{S}$ with $k=1,2,..,S$.
For example, by varying $\gamma$, edges start splitting when $\gamma<\frac{k}{S}$. 
Hence, at the critical values $\frac{k}{S}$ with $k=1,2,..,S$, we have a different number of edges susceptible to influence or split and different final magnetization.

The final result is that the boundaries of the bands for sparse hypergraphs is given by the following theorem.
\begin{theorem}
Let an $S$-uniform hypergraph with $S>2$, $N$ nodes, $n$ edges, $p$ the probability of rewiring, $\gamma$ the threshold parameter evolve following the dynamics described in Sect.~\ref{sec:model}, without the merging and rewiring mechanism.
For low mean degrees of the initial hypergraph (e.g., $\frac{Sn}{N}\approx 1$), the magnetization discontinuously changes at the following critical values of $\gamma$
\begin{equation}
    \gamma_c=\frac{k}{S},
\end{equation}
where $k\in \mathbb{Z}$ such that $N(k,S,0)>0$ and $\frac{k}{S}\leq\frac{1}{2}$ or:
\begin{equation}
    \gamma_c=1-\frac{k}{S},
\end{equation}
where $k\in \mathbb{Z}$ such that $N(k,S,0)>0$ and $\frac{k}{S}>\frac{1}{2}$
\label{theorem_bands}
\end{theorem}

\begin{figure}[htb]
\begin{subfigure}[b]{0.5\textwidth}
    \centering
  \includegraphics[width=\linewidth]{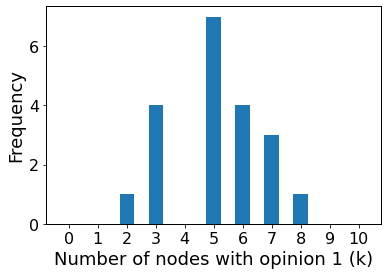}
  \caption{Histograms of $N(k,10,0)$}
  \label{histogram_nklt}
\end{subfigure}
\hfill
\begin{subfigure}[b]{0.5\textwidth}
    \centering
  \includegraphics[width=\linewidth]{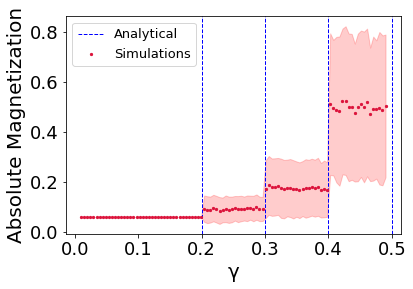}
  \caption{Absolute magnetization vs $S\gamma$}
  \label{m_vs_sgamma}
\end{subfigure}
\caption{Illustration of Theorem \ref{theorem_bands}, which calculates the locations of the bands. Parameters: $N=100$, $p=0.5$, $S=10$, $n=20$ edges. Subfigure \ref{histogram_nklt} shows the frequency of $N(k,10,0)$ for $k\in\{0,..,10\}$ where $N(k,l,t)$ is the number of edges of size $l$ with $k$ nodes with opinion 1 at time $t$. The values of $k$ with non-zero $N(k,10,0)$ are $k=\{2,3,5,6,7,8\}$. Based on Theorem \ref{theorem_bands}, we calculate the minority fraction of the edges $(k,10)$ with $k=\{2,3,5,6,7,8\}$ and this gives the critical values of the bands $\gamma_c=\{\frac{2}{10},\frac{3}{10},\frac{4}{10},\frac{5}{10}\}$ which match with the simulations in Subfigure \ref{m_vs_sgamma}. In Subfigure \ref{m_vs_sgamma}, the blue vertical lines are the positions of the bands calculated by the previous theorem.}
\label{fig:comparison_theorem_uniform}
\end{figure}

In Figure \ref{fig:comparison_theorem_uniform}, we show that the expectations coming from analytical analysis match the simulation results for mean degree equal to $2$. 
For systems with high mean degrees (approximately higher than $4$), the previous analysis does not work because inactive edges can still get re-activated due to overlap.
However, the boundaries of the bands still occur at rational numbers $\frac{k}{S}$ of $\gamma$.
We argue that this occurs as the trajectories in which inactive edges get re-activated are rare and do not contribute significantly to the final state.

\subsubsection{The height of the bands}
We calculate the height of the bands, i.e., the final absolute magnetization at equilibrium for sparse hypergraphs. 
To calculate this, we first characterize initial configurations depending on their fraction of edges having a certain majority.
Then, we use this information to calculate the probability of different initial configurations.
Finally, we compute the final expected magnetization by computing the expected final state of each hyperedge based on its initial majority.
The expected final state of each edge is obtained under the sparse hypergraph assumption.

To estimate the initial expected magnetization, recall that given an initial $A_s$ there are many different possible initial configurations.
For example, if $A_s=0.55$, we can have with a high probability that about half of the nodes have opinion 1; and with low with probability, we can also have that all the nodes have opinion 0.
To account for these different initial configurations, we consider the following binomial probability for observing $k$ nodes with opinion 1:
\begin{equation}
\label{eq:alphas}
    p(k, A_s) = {{N}\choose{k}} A_s^{k}(1-A_s)^{N-k}
\end{equation}
Then, the probability to observe an edge of size $S$ with $\lambda$ nodes with opinion 1 at fixed initial fraction $\alpha=k/N$ is:\footnote{To write this last equation, we assume that the system is large enough as we are using a sampling with replacement.
The exact formula is instead: $  q(\lambda, k, N) = \frac{ {{k}\choose{\lambda} } {{N-k}\choose{S-\lambda}}} {  {{S}\choose{N}}}$}
\begin{equation}
\label{eq:hyperedge_densities_at_fixed_alpha}
    p(\lambda, k/N) = n\cdot{{S}\choose{\lambda} }\left(\frac{k}{N}\right)^{\lambda}\left(1-\frac{k}{N}\right)^{S- \lambda}
\end{equation}
By taking the product of $p(\lambda, k/N)$ and $p(k, A_s)$, we obtain the probability to observe a edge with $\lambda$ nodes with opinion 1 in an initial configuration with $k$ nodes with opinion 1.
Using this probability, we can compute the initial expected magnetization:
\begin{equation}
\label{eq:sparse_init_mag}
    \langle m(0, A_s) \rangle = \sum_{k=0}^{N}\sum_{\lambda=0}^{\min{(S,k)}}p(k,A_s) p(\lambda, k/N) n \cdot  m(\lambda)
\end{equation}
where $m(\lambda) = \frac{1}{Sn}\left(2\lambda-S\right)$ is the magnetization of an edge with $\lambda$ nodes with opinion 1 and $n$ is the number of hyper-edges.

From \eqref{eq:sparse_init_mag}, we obtain the final expected magnetization by recalling that when a hypergraph is sparse, the evolution of its edges are independent and hence, determined by the magnitude of the initial majority:
\begin{enumerate}
    \item if $\gamma<\frac{\lambda}{S}<1-\gamma$ then $e$ splits and the final magnetization does not change,
    \item if $\frac{\lambda}{S}>1-\gamma$ then all its nodes with opinion $0$ will eventually get opinion $1$,
    \item if $\frac{\lambda}{S}<\gamma$ then all its nodes with opinion $1$ will eventually get opinion $0$.
\end{enumerate}
By applying, these three conditions to \eqref{eq:sparse_init_mag}, we compute the expected final magnetization:
\begin{equation}
\label{eq:sparse_height}
\begin{split}
    \langle m(\infty, A_s)\rangle = \sum_{k=0}^{N}p(k,A_s)&\left[\quad
    \sum_{\lambda=0}^{S\gamma}p(\lambda, k/N)m(\lambda) +\right.\\
    &+\sum_{\lambda=S\gamma}^{S-S\gamma}p(\lambda, k/N)m(\lambda)+\\
    &+\left.\sum_{\lambda=S-S\gamma}^Sp(\lambda, k/N)m(\lambda)
    \right]n
\end{split}
\end{equation}

\begin{figure}[htb]
\begin{subfigure}[b]{0.5\textwidth}
    \centering
  \includegraphics[width=\linewidth]{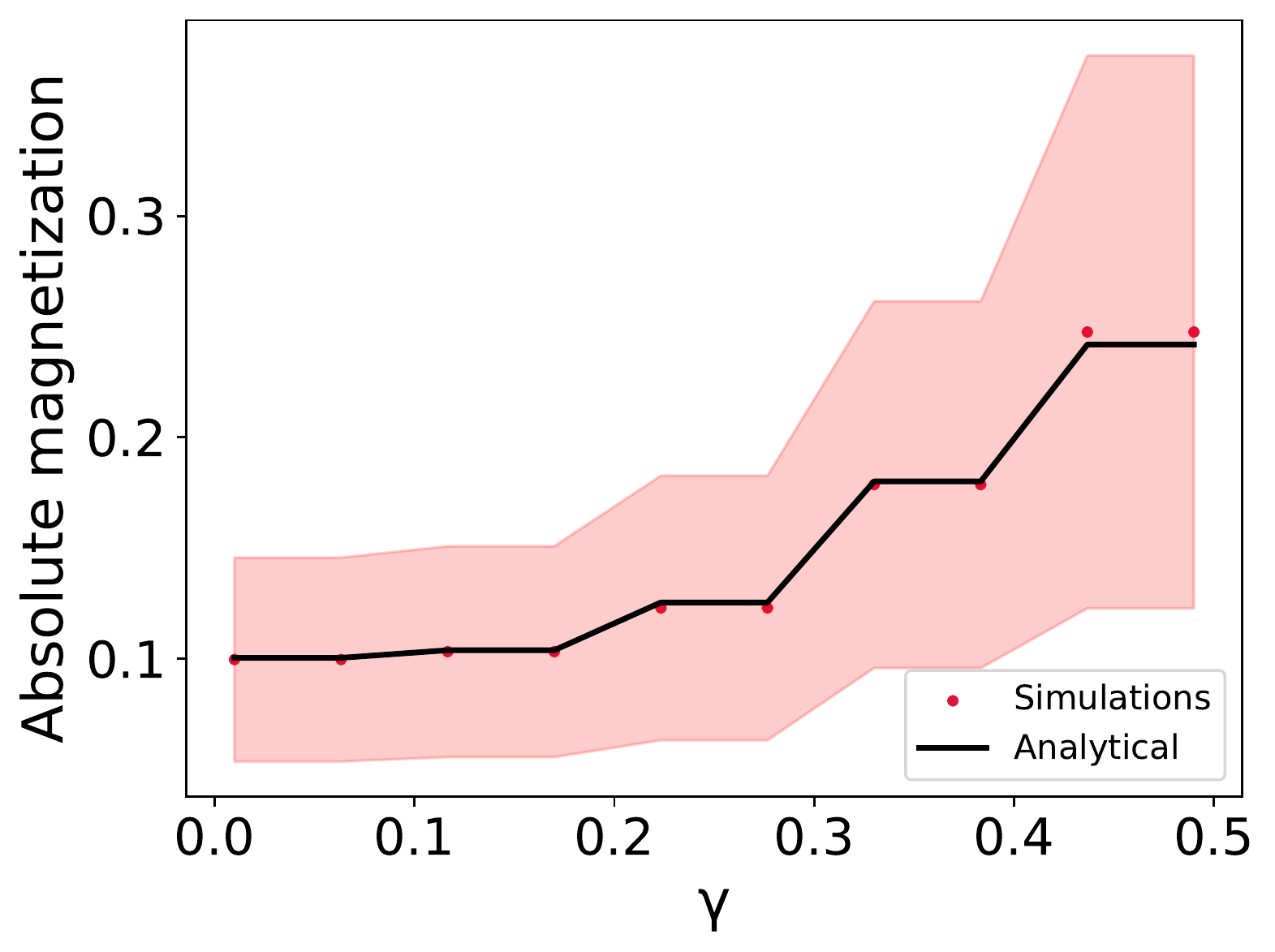}
  \caption{Mean Degree 1}
  \label{sparse_degree_1}
\end{subfigure}
\hfill
\begin{subfigure}[b]{0.5\textwidth}
    \centering
  \includegraphics[width=\linewidth]{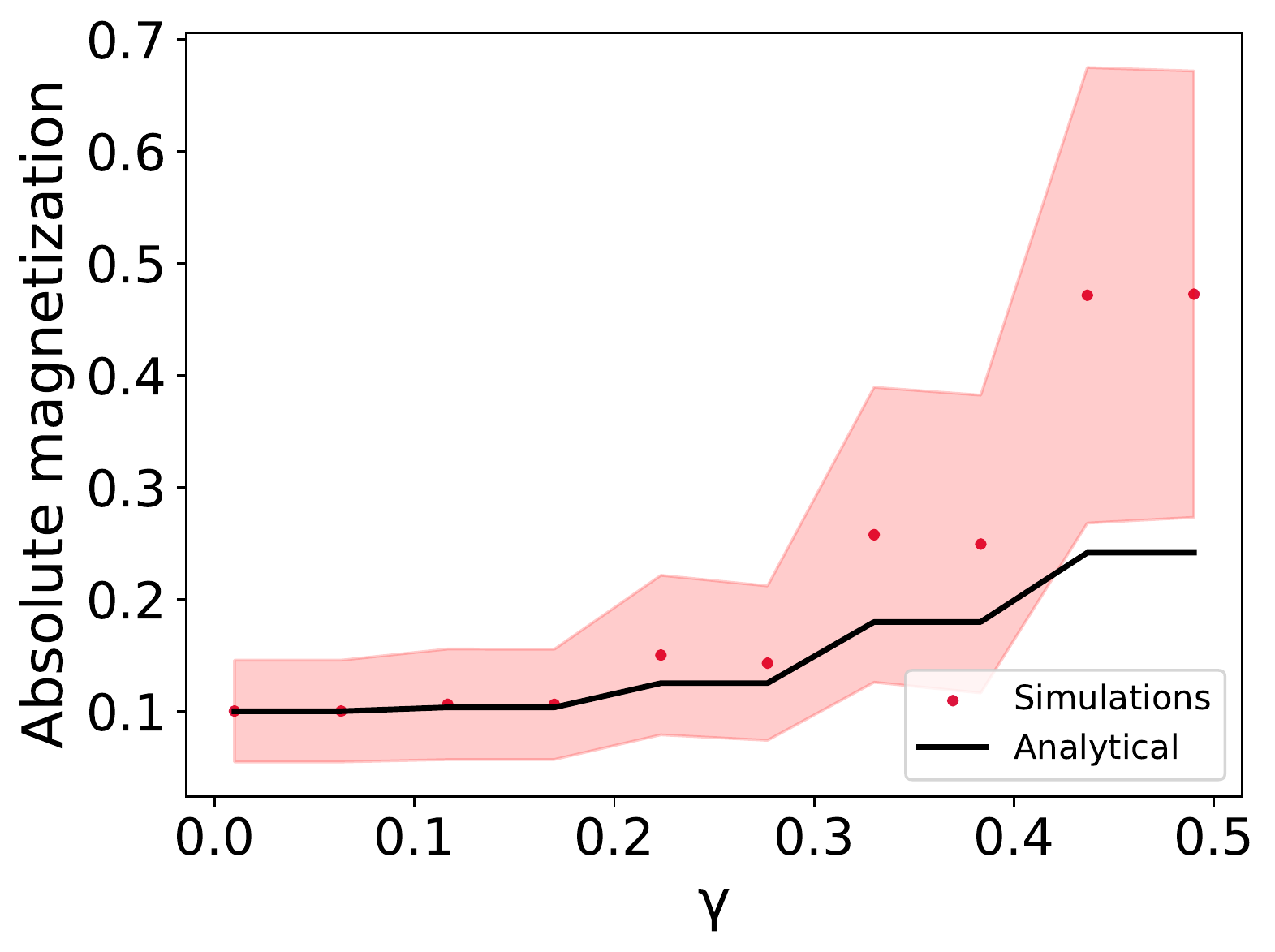}
  \caption{Mean Degree 1.5}
  \label{sparse_degree_15}
\end{subfigure}
\caption{Comparison of the Analysis with the Sparsity Approximation (Equation \ref{eq:sparse_height}) and the simulations for mean degree 1 and 1.5. Parameters: $p=0.55$, $N=500$, $S=10$, $n=50$ (for Subfigure \ref{sparse_degree_1}), $n=75$ (for Subfigure \ref{sparse_degree_15}), $A_S=0.55$. The points and the shaded area are the mean and the standard deviation of the trajectories respectively for 1000 initial configurations. The black line is the analytical magnetization based on equation \ref{eq:sparse_height}.}
\label{fig:sparsity_comparison}
\end{figure}
\begin{figure}[htb]
\centering
  \includegraphics[width=0.75\textwidth]{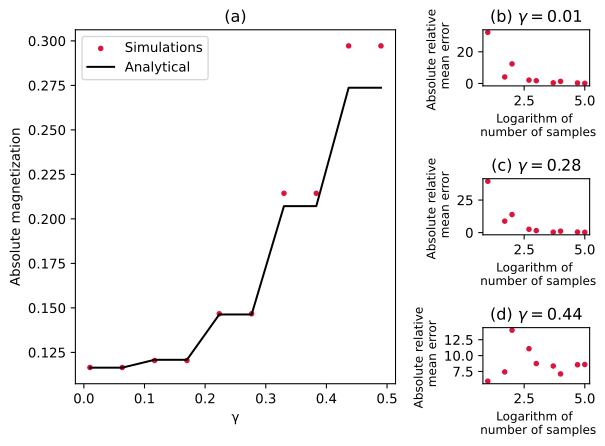}
\caption{Comparison of the Analysis with the Sparsity Approximation (Equation \ref{eq:sparse_height}) and the simulations for mean degree 1. Parameters: $p=0.55$, $N=100$, $S=10$, $n=10$, $A_S=0.55$. The points are the mean of the trajectories respectively for XXX trajectories. On the right, we have the convergence between the simulations and the analytical absolute magnetization when increasing the sample size.}
\label{fig:sparsity_comparison_mean_degree1}
\end{figure}

In Figure~\ref{fig:sparsity_comparison}, we compare \eqref{eq:sparse_height} with the simulations for mean degree equal to $1$ and $1,5$.
The analytic predictions for the final absolute magnetization are compatible with the values coming from the simulations.
We observe that the match is better for mean degree equal to $1$.
This is expected as when the mean degree is low, then the sparsity assumption is less violated. 
In Fig.\ref{fig:sparsity_comparison_mean_degree1}, we have perfect match between the analytic formula and the simulations for the limit case of mean degree equal to 1.
Precisely, we obtain a relative error lower than 10\% which decreases with increasing sample size (see left panels in Fig.\ref{fig:sparsity_comparison_mean_degree1}).
Also, note that the relative error is higher for larger $\gamma$.
This possibly occurs because the number of possible final states increases with $\gamma$ and hence, we require more simulations to explore them. 

When increasing the mean degree, the analytic predictions based on \eqref{eq:sparse_height} significantly underestimates the simulated values. 
The reason is that the analytic prediction is valid under the assumption that there is no overlap between edges.
For high mean degree, the overlap is instead significant and allows the initial global majority opinion to better diffuse in the system.
Precisely, recall that it is more likely to sample a hyperedge whose local majority is equal to the global majority.
In this sampled edge, nodes might change their opinion to the global majority opinion.
Because of the overlap, these changes increase the number of nodes having the global majority opinion not only in the sampled edge but also in its overlapping edges.
Thus, the expected number of edges with the local majority opinion equal to the global majority opinion increases.
In other words, overlap increases the absolute final magnetization.

\section{Conclusion}
\label{sec:conclusion}

We have extended the adaptive voter model by including group interactions.
We have shown that the inclusion of group interactions drastically changes the dynamics and can lead to fragmentation bands at equilibrium.
Specifically, fragmentation bands appear at equilibrium because the final global majority can reach different values. 
This final value depends on the critical size a majority should have in a group to convince the minority to change their opinion. 
This type of final state is not present with only pairwise interactions.
Note that this type of final state is not present in the absence of group interactions.

Second, different groups may share individuals, and this group overlap might create geometrical frustration. 
The presence of this frustration creates $k$-orbits as final states. 
Therefore, unlike the classical adaptive voter model, the system does not always reach a single equilibrium configuration. 
Instead, the system may get trapped in oscillations where some individuals change their opinions periodically. 
Although this finding might not have direct applications, it shows how group interactions enrich the set of possible final states from a mathematical point of view.

For analytical tractability, we have assumed an $S$-uniform hypergraph as the initial topology. 
This is a simplification since real-world social hypergraphs are instead heterogeneous as individuals interact in groups of different sizes and with non-random connectivity. 
Recently, in \citep{landry2020effect, landry2023polarization}, it was shown that the heterogeneity of the initial configuration in presence higher order interactions can significantly affect the system dynamics.
Based on our results, we expect instead no drastic changes in the dynamics, but only an increase in the number of fragmentation bands.
It is still open for research to study the interplay between group interactions and heterogeneous initial configurations.

Finally, extending the model to consider individuals with different importance would be interesting. 
Some individuals might be more influential because of their status\citep{pilgrim2020organisational}, reputation\citep{jain2022trust}, or position in a hierarchical structure\citep{o2021hierarchical}. 
It would be straightforward to account for the importance of each individual when computing group majorities. 
This extension would allow for modelling scenarios where a more \textit{silent majority} adopts the opinion of a \textit{louder minority}.

\section*{Acknowledgements }
The authors thank Frank Schweitzer and Luca Verginer for providing useful suggestions about the notation and visualizations.

\bibliography{refs} 

\begin{thebibliography}{10}

\bibitem{brush1967history}
S.~G. Brush, ``History of the lenz-ising model,'' {\em Reviews of modern
  physics}, vol.~39, no.~4, p.~883, 1967.

\bibitem{adaptivevotermodel}
R.~Durrett, J.~P. Gleeson, A.~L. Lloyd, P.~J. Mucha, F.~Shi, D.~Sivakoff, J.~E.
  Socolar, and C.~Varghese, ``Graph fission in an evolving voter model,'' {\em
  Proceedings of the National Academy of Sciences}, vol.~109, no.~10,
  pp.~3682--3687, 2012.

\bibitem{iacopini2019simplicial}
I.~Iacopini, G.~Petri, A.~Barrat, and V.~Latora, ``Simplicial models of social
  contagion,'' {\em Nature communications}, vol.~10, no.~1, p.~2485, 2019.

\bibitem{iacopini2022group}
I.~Iacopini, G.~Petri, A.~Baronchelli, and A.~Barrat, ``Group interactions
  modulate critical mass dynamics in social convention,'' {\em Communications
  Physics}, vol.~5, no.~1, p.~64, 2022.

\bibitem{landry2023polarization}
N.~W. Landry and J.~G. Restrepo, ``Polarization in hypergraphs with community
  structure,'' {\em arXiv preprint arXiv:2302.13967}, 2023.

\bibitem{ferraz2023multistability}
G.~Ferraz~de Arruda, G.~Petri, P.~M. Rodriguez, and Y.~Moreno,
  ``Multistability, intermittency, and hybrid transitions in social contagion
  models on hypergraphs,'' {\em Nature Communications}, vol.~14, no.~1,
  p.~1375, 2023.

\bibitem{brown1986perceptions}
B.~B. Brown, D.~R. Clasen, and S.~A. Eicher, ``Perceptions of peer pressure,
  peer conformity dispositions, and self-reported behavior among
  adolescents.,'' {\em Developmental psychology}, vol.~22, no.~4, p.~521, 1986.

\bibitem{milgram1969note}
S.~Milgram, L.~Bickman, and L.~Berkowitz, ``Note on the drawing power of crowds
  of different size.,'' {\em Journal of personality and social psychology},
  vol.~13, no.~2, p.~79, 1969.

\bibitem{latane1981social}
B.~Latan{\'e} and S.~Wolf, ``The social impact of majorities and minorities.,''
  {\em Psychological Review}, vol.~88, no.~5, p.~438, 1981.

\bibitem{holyst2000phase}
J.~A. Ho{\l}yst, K.~Kacperski, and F.~Schweitzer, ``Phase transitions in social
  impact models of opinion formation,'' {\em Physica A: Statistical Mechanics
  and its Applications}, vol.~285, no.~1-2, pp.~199--210, 2000.

\bibitem{goban2018emergence}
A.~Goban, R.~Hutson, G.~Marti, S.~Campbell, M.~Perlin, P.~Julienne,
  J.~D’incao, A.~Rey, and J.~Ye, ``Emergence of multi-body interactions in a
  fermionic lattice clock,'' {\em Nature}, vol.~563, no.~7731, pp.~369--373,
  2018.

\bibitem{pelka2020chimera}
K.~Pelka, V.~Peano, and A.~Xuereb, ``Chimera states in small optomechanical
  arrays,'' {\em Physical Review Research}, vol.~2, no.~1, p.~013201, 2020.

\bibitem{yu2011higher}
S.~Yu, H.~Yang, H.~Nakahara, G.~S. Santos, D.~Nikoli{\'c}, and D.~Plenz,
  ``Higher-order interactions characterized in cortical activity,'' {\em
  Journal of neuroscience}, vol.~31, no.~48, pp.~17514--17526, 2011.

\bibitem{levine2017beyond}
J.~M. Levine, J.~Bascompte, P.~B. Adler, and S.~Allesina, ``Beyond pairwise
  mechanisms of species coexistence in complex communities,'' {\em Nature},
  vol.~546, no.~7656, pp.~56--64, 2017.

\bibitem{battiston2020networks}
F.~Battiston, G.~Cencetti, I.~Iacopini, V.~Latora, M.~Lucas, A.~Patania, J.-G.
  Young, and G.~Petri, ``Networks beyond pairwise interactions: structure and
  dynamics,'' {\em Physics Reports}, 2020.

\bibitem{klimm2021hypergraphs}
F.~Klimm, C.~M. Deane, and G.~Reinert, ``Hypergraphs for predicting essential
  genes using multiprotein complex data,'' {\em Journal of Complex Networks},
  vol.~9, no.~2, p.~cnaa028, 2021.

\bibitem{bairey2016high}
E.~Bairey, E.~D. Kelsic, and R.~Kishony, ``High-order species interactions
  shape ecosystem diversity,'' {\em Nature communications}, vol.~7, no.~1,
  p.~12285, 2016.

\bibitem{golubski2016ecological}
A.~J. Golubski, E.~E. Westlund, J.~Vandermeer, and M.~Pascual, ``Ecological
  networks over the edge: hypergraph trait-mediated indirect interaction (tmii)
  structure,'' {\em Trends in ecology \& evolution}, vol.~31, no.~5,
  pp.~344--354, 2016.

\bibitem{taramasco2010academic}
C.~Taramasco, J.-P. Cointet, and C.~Roth, ``Academic team formation as evolving
  hypergraphs,'' {\em Scientometrics}, vol.~85, no.~3, pp.~721--740, 2010.

\bibitem{salnikov2018simplicial}
V.~Salnikov, D.~Cassese, and R.~Lambiotte, ``Simplicial complexes and complex
  systems,'' {\em European Journal of Physics}, vol.~40, no.~1, p.~014001,
  2018.

\bibitem{bick2021higher}
C.~Bick, E.~Gross, H.~A. Harrington, and M.~T. Schaub, ``What are higher-order
  networks?,'' {\em arXiv preprint arXiv:2104.11329}, 2021.

\bibitem{nk2022}
N.~Papanikolaou, G.~Vaccario, E.~Hormann, R.~Lambiotte, and F.~Schweitzer,
  ``Consensus from group interactions: An adaptive voter model on
  hypergraphs,'' {\em Physical Review E}, vol.~105, no.~5, p.~054307, 2022.

\bibitem{noonan2021dynamics}
J.~Noonan and R.~Lambiotte, ``Dynamics of majority rule on hypergraphs,'' {\em
  arXiv preprint arXiv:2101.03632}, 2021.

\bibitem{landry2020effect}
N.~W. Landry and J.~G. Restrepo, ``The effect of heterogeneity on hypergraph
  contagion models,'' {\em Chaos: An Interdisciplinary Journal of Nonlinear
  Science}, vol.~30, no.~10, p.~103117, 2020.

\bibitem{pilgrim2020organisational}
C.~Pilgrim, W.~Guo, and S.~Johnson, ``Organisational social influence on
  directed hierarchical graphs, from tyranny to anarchy,'' {\em Scientific
  Reports}, vol.~10, no.~1, pp.~1--13, 2020.

\bibitem{jain2022trust}
E.~Jain and A.~Singh, ``Trust-and reputation-based opinion dynamics modelling
  over temporal networks,'' {\em Journal of Complex Networks}, vol.~10, no.~4,
  p.~cnac019, 2022.

\bibitem{o2021hierarchical}
J.~D. O'Brien, K.~A. Oliveira, J.~P. Gleeson, and M.~Asllani, ``Hierarchical
  route to the emergence of leader nodes in real-world networks,'' {\em
  Physical Review Research}, vol.~3, no.~2, p.~023117, 2021.

\end{thebibliography}
\bibliographystyle{ieeetr}

\appendix
\section{Proof of Location of Bands}
\label{proof_appendix}
We prove Theorem \ref{theorem_bands}, which describes the locations of the bands for $S$-uniform initial configurations.

Let's assume a system with $N$ nodes initialized as an $S$-uniform hypergraph $\mathcal{H}$ at $t=0$ with $n$ edges. The threshold parameter is $\gamma$ and the probability of rewiring is $p$. The quantity $N(k,l,t)$ describes the number of edges in the hypergraph with $k$ nodes with opinion 1 and size $l$ at time $t$.

At a time $t$ a random edge is selected and the dynamics of Section \ref{sec:model} are applied to it. To make the problem analytically tractable we switch off merging/rewiring. We also assume that at each timestep only one node of the selected edge flips its opinion. This assumption highly simplifies the analysis. It is also reasonable since the case where e.g. $\nu$ nodes change opinions at a timestep occurs with probability $\pi^\nu$  where $\pi$ is the probability that each minority in the selected edge changes its opinion. $\pi^\nu$ is a small number since $\pi\leq0.5$ because it depends on the minority fraction. We list all the possible changes to  $N(k,l,t)$ for some $k$ and $l$ after the dynamics act on the specific edge selected. To start, we ignore changes to $N(k,l,t)$ due to overlap of edges with the selected edge but we consider them later on. 

\begin{equation}
        N(k,l,t+1)= 
\begin{cases}
    N(k,l,t)+1,& \text{with $p_1$ if } \frac{k-1}{l}> 1-\gamma  \text{ and $(k-1,l)$ selected},\\
    N(k,l,t)+1,& \text{with $p_2$ if } \frac{k+1}{l} < \gamma  \text{ and $(k+1,l)$ selected},\\
    N(k,l,t),& \text{with $1-p_1$ if } \frac{k-1}{l}> 1-\gamma  \text{ and $(k-1,l)$ selected},\\
    N(k,l,t),& \text{with $1-p_2$ if } \frac{k+1}{l} < \gamma  \text{ and $(k+1,l)$ selected},\\
    N(k,l,t)-1,& \text{with $p_3$ if } \frac{k}{l} > 1-\gamma  \text{ and $(k,l)$ selected},\\
    N(k,l,t)-1,& \text{with $p_4$ if } \frac{k}{l} < \gamma  \text{ and $(k,l)$ selected},\\
    N(k,l,t),& \text{with $1-p_3$ if } \frac{k}{l} > 1-\gamma  \text{ and $(k,l)$ selected},\\
    N(k,l,t),& \text{with $1-p_4$ if } \frac{k}{l} < \gamma  \text{ and $(k,l)$ selected},\\
    N(k,l,t),& \text{if } \frac{k-1}{l} \leq 1-\gamma  \text{ and $(k-1,l)$ selected},\\
    N(k,l,t),& \text{if } \frac{k+1}{l} \geq \gamma  \text{ and $(k+1,l)$ selected},\\
    N(k,l,t)-1,& \text{if } \gamma \leq \frac{k}{l} \leq 1-\gamma  \text{ and $(k,l)$ selected},\\
    N(k,l,t),& \text{otherwise }
\end{cases},
\label{master_k_uniform}
\end{equation}
where $p_1, p_2, p_3, p_4$ are transition probabilities that need to be found and "$(k,l)$ selected" means the edge selected has $k$ nodes of opinion 1 and size $l$. For the quantities $N(k,k,t)$ and $N(0,l-k,t)$ there are two extra cases resulting from the splitting process:
\begin{equation}
\begin{split}
N(k,k,t+1)&=
\begin{cases}
    &... , \\ N(k,k,t)+1,&\text{if }  \gamma \leq \frac{k}{l} \leq 1-\gamma  \text{ and $(k,l)$ selected}\ ,\\
\end{cases}
\end{split}
\label{exception1}
\end{equation}
and \begin{equation}
\begin{split}
N(0,l-k,t+1)&= 
\begin{cases}
    &... , \\N(0,l-k,t)+1,& \text{if }  \gamma \leq \frac{k}{l} \leq 1-\gamma  \text{ and $(k,l)$ selected}\ \\
\end{cases}
\end{split}
\label{exception2}
\end{equation}

The intuition behind these lists is the following: $N(k,l,t)$ can change only if the edge selected at the given timestep is a $(k-1,l)$, a $(k+1,l)$ or a $(k,l)$ edge since we assumed merge and rewire is switched off and only one node changes opinion at each timestep. A small exception to this rule is when we study $N(0,k,t)$ or $N(k,k,t)$  since we also need to consider an increase because of the split edges created by the split mechanism (Equations \ref{exception1} and \ref{exception2}). In Equation \ref{master_k_uniform}, the first two cases depending on $p_1$ and $p_2$ consider increase of $N(k,l,t)$ because influence occurs on $(k-1,l)$ and $(k+1,l)$ edges respectively changing the opinion of a node from 0 to 1 and 1 to 0 respectively. In the 6th and 7th cases depending on $p_3$ and $p_4$ we consider the possibility that $N(k,l,t)$ decreases when $(k,l)$ is selected and influence occurs changing a node from 0 to 1 or 1 to 0 respectively. The second to last case considers the possibility that $N(k,l,t)$ decreases because a $(k,l)$ edge splits. 

However, the transition probabilities have not yet been defined. These depend on the probability of selecting an edge of a particular type e.g. $(k,l)$ and on the probability that influence or splitting occurs. In addition, we have to consider changes to $N(k,l,t)$ due to overlap of the edges further increasing the correlations between the edges. To make this analysis tractable we use a mean field approximation. We apply the following assumptions: 

\begin{itemize}
    \item the probability a node $i$ in a selected $(k,l)$ edge $e$ has opinion $s_i(t)$ at time $t$ is equal to the fraction of nodes with opinions $s_i(t)$ in $(k,l)$:
    \begin{equation}
        \begin{aligned}
            p(s_i=0\mid i\in e \text{ where $e$ is a selected $(k,l)$ edge})=\frac{l-k}{l} \\
            p(s_i=1\mid i\in e \text{ where $e$ is a selected $(k,l)$ edge})=\frac{k}{l}
        \end{aligned}
    \end{equation}
    \item given that a node $j$ in a selected edge $(k,l)$ changes its opinion $s_j(t)=0$ to $s_j(t+1)=1$ ($s_j(t)=1$ to $s_j(t+1)=0$), a node $i$ with opinion $0$ ($1$) in the system also changes its opinion with a probability proportional $l-k$ ($k$): 
    \begin{equation}
        \begin{aligned}
            p\Bigg(s_i(t+1)=1 \Bigg| \begin{matrix}
s_i(t)=0, s_j(t)=0, s_j(t+1)=1, j\in e \\ \text{ where $e$ is a selected $(k,l)$ edge})
    
\end{matrix} \Bigg) =b (l-k) \\
            p\Bigg(s_i(t+1)=0 \Bigg| \begin{matrix}
s_i(t)=1, s_j(t)=1, s_j(t+1)=0, j\in e \\ \text{ where $e$ is a selected $(k,l)$ edge})

\end{matrix} \Bigg) =b k
        \end{aligned},
    \end{equation}    
    where $b$ is the proportionality constant. 
\end{itemize}

Combining these probabilities, the probability that a node with opinion $0$ or $1$ in a selected $(k,l)$ edge $e$ changes its opinion is respectively: 
\begin{equation}
        \begin{aligned}
            p(s_i(t+1)=1\mid, s_i(t)=0, i\in e \text{ where $e$ is a selected $(k,l)$ edge})=\frac{b(l-k)^2}{l} \\
            p(s_i(t+1)=0\mid, s_i(t)=1, i\in e \text{ where $e$ is a selected $(k,l)$ edge})=\frac{bk^2}{l}
    \end{aligned},
    \label{transprob}
\end{equation}

We can apply these assumptions for overlapping edges as well and  we can now express the transition probabilities as $p_1=\frac{b(l-k+1)^2}{l}, p_2=\frac{b(k+1)^2}{l}, p_3=\frac{b(l-k)^2}{l}, p_4=\frac{bk^2}{l}$ for a selected edge $(k,l)$. 

To study the bands, we assume without loss of generality broken symmetry with positive initial magnetization. That means we assign opinion 1 to nodes more frequently than we assign opinion 0 to the initial configuration. We fix $k=l$ and study the change of $N(l,l,t)$ at one timestep:
\begin{equation}
    \langle \delta N(l,l,t+1) \rangle =  \langle N(l,l,t+1) \rangle - N(l,l,t).
\end{equation}
When $\langle \delta N(l,l,t+1) \rangle>0$ then there are $(l,l)$ edges which contribute long term to magnetization. They contribute long term because they are inactive and therefore they are probabilistically fixed. "Probabilistically" because, due to overlap, some nodes of those fixed edges can change their opinion re-activating the inactive edges. Therefore, for the rest of the analysis we assume that the overlap is low enough (i.e. low mean degree) such that inactive edges on average stay inactive.

Let $p(1,k,l,t+1)$ be the probability that $N(k,l,t)$ increases by $1$ at $t$. Similarly, we also define $p(0,k,l,t+1)$ and $p(-1,k,l,t+1)$. 

Using $p_1, p_2, p_3, p_4$ and Equation \ref{master_k_uniform}, $\langle \delta N(k,l,t+1) \rangle$ can be expressed as:
\begin{equation}
    \begin{split}
        \langle \delta N(k,l,t+1) \rangle =\frac{1}{3}\Bigg( \frac{b(l-k+1)^2}{l}H(\frac{k-1}{l}-1+\gamma)p(k-1,l,t-1)+\\+\frac{b(1+k)^2}{l}H(\gamma-\frac{k+1}{l})p(k+1,l,t-1)-\\-\bigg(\frac{b(l-k)^2}{l}H(\frac{k}{l}-1+\gamma)+\frac{bk^2}{l}H(\gamma-\frac{k}{l})+\frac{bk^2}{l}H(\frac{k}{l}-\gamma)H(1-\gamma-\frac{k}{l})\bigg)p(k,l,t-1)\Bigg),
        \end{split}
    \label{general_domino}
\end{equation}
where $p(k,l,t)$ is the probability of selecting an edge $(k,l)$ at time $t$ and $H(x)$ is the Heaviside step function ($H(x)=1$ for $x>0$, else $H(x)=0$).

At $t=0$ the hypergraph is $S$-uniform with $n$ edges and therefore $N(k,S,0)=n$ for $k\in\{0,1,...,S\}$. We derive the value of $\gamma$ for which $\langle \delta N(S,S,t+1) \rangle>0$ at a timestep $t$, since we expect the edges $(S,S)$ to contribute the most to the magnetization because they have the largest number of nodes with opinion 1 and they are fixed. 

Using Equation \ref{general_domino} for $(S,S)$ edges we get:
\begin{equation}
    \langle \delta N(S,S,t+1) \rangle = \frac{b}{3l}H(\gamma-\frac{1}{S})p(S-1,S,t).
    \label{domino}
\end{equation}
We try to find a condition of the threshold parameter $\gamma$ such that there exists a $t$ when $\langle \delta N(S,S,t+1) \rangle >0$. If there is not such $t$ then the quantity $\langle \delta N(S,S,t+1) \rangle $ is always $0$ and we do not expect to see clear characteristic bands. Therefore, we assume there exists $t$ when $\langle \delta N(S,S,t+1) \rangle >0$. Let $\tau$ be the minimum time where $\langle \delta N(S,S,\tau) \rangle >0$. This means that $\gamma > \frac{1}{S}$ and $p(S-1,S,\tau)>0$. If $p(S-1,S,\tau)>0$ then there exists at least one edge $(S-1,S)$ at time $\tau$ i.e. $N(S-1,S,\tau)>0$. Due to our initial conditions, this means either that $N(S-1,S,0)>0$  or that there is a $\tau'<\tau$ such that $\langle \delta N(S-1,S,\tau') \rangle >0$. The latter imposes additional conditions since, using Equation \ref{general_domino},  $\langle \delta N(S-1,S,\tau') \rangle$ can be expressed as:
\begin{equation}
    \begin{split}
        \langle \delta N(S-2,S,\tau') \rangle = \frac{1}{3}\Bigg(\frac{4b}{S}H(-\frac{2}{S}+\gamma)p(S-2,S,\tau'-1)\\-\bigg(\frac{b}{S}H(\gamma-\frac{1}{S})+\frac{b(S-1)^2}{S}H(\gamma-1+\frac{1}{S})\\+\frac{b(S-1)^2}{S}H(-\gamma+1-\frac{1}{S})H(\frac{1}{S}-\gamma)\bigg)p(S-1,S,\tau'-1)\Bigg)
    \end{split}
    \label{domino2}
\end{equation}

Thus, $\langle \delta N(S-2,S,\tau') \rangle>0$ is positive if $\gamma >\frac{2}{S}$ and $p(S-2,S,\tau'-1)>0$. If  $\gamma \leq \frac{2}{S}$ there is not $\tau'<\tau$ such that $\langle \delta N(S-2,S,\tau') \rangle>0$ at the expense of $N(S,S,t)$ and the magnetization. If $\gamma >\frac{2}{S}$ and $p(S-2,S,\tau'-1)>0$ we repeat the previous procedure. This creates a domino effect until $t=0$. Therefore, the significant contributions to the magnetization depends on the initial edges $N(k,S,0)$ where  $k\in\{\ceil{\frac{S}{2}},\ceil{\frac{S}{2}}+1,...,S-1,S\}$. This means that at $t=1$, the initial edges that do not split, i.e. $\gamma \geq\frac{k}{S}$  such that $N(k,S,0)>0$,  become $(S,S)$ edges contributing to $N(S,S,t)$. Increasing $\gamma$ leads to edges $(k,S)$ with lower value of $k$ at $t=0$ not splitting and becoming inactive $(S,S)$. 

Similarly, we can follow the same procedure for negative initial magnetizations by studying $(0,S)$ edges.  $\square$ %

\end{document}